\documentclass{llncs}
\pdfoutput=1

\usepackage{hyperref}
\usepackage{etex}
\usepackage{color}
\usepackage[usenames,dvipsnames]{xcolor}
\usepackage{listings, multicol}	
\usepackage{enumerate}
\usepackage{layouts}
\usepackage{times}
\usepackage{natbib}
\usepackage{notoccite}
\usepackage{todonotes}
\usepackage{alltt}
\usepackage{url}
\usepackage{amsmath}
\usepackage{amsfonts}
\usepackage{caption}
\usepackage{paralist}
\usepackage{soul}
\usepackage{enumitem}
\usepackage{appendix}

\usepackage{fancyhdr}
\usepackage{datetime}

\usepackage{amsthm}

\usepackage{xspace}
\usepackage{tikz}
\usepackage{subcaption}
\usepackage{calc}
\usetikzlibrary{calc}

\usepackage{fancyvrb}

\theoremstyle{definition}
\newcommand{\triple}[1]{\ensuremath{\langle #1 \rangle}}
\newcommand{\pair}[1]{\ensuremath{\left(#1\right)}}
\newcommand{\graph}[1]{\ensuremath{\mathcal{#1}}}
\newcommand{\graphset}[1]{\ensuremath{\mathbb{#1}}}

\newcommand{\NP}{\ensuremath{\mathbf{NP}}}
\newcommand{\coNP}{\ensuremath{\mathbf{coNP}}}
\newcommand{\QBF}{\ensuremath{\mathbf{QBF}}}
\newcommand{\theory}{\ensuremath{\mathcal{T}}}

\lstdefinestyle{model}{
  mathescape,
  columns=fullflexible,
  numbers=left,
  frame=single,
  escapechar=@,
  basicstyle=\scriptsize\ttfamily
}

\lstdefinestyle{small}{
  basicstyle=\scriptsize\ttfamily,
  breaklines=true
}

\RecustomVerbatimCommand{\VerbatimInput}{VerbatimInput}%
{fontsize=\footnotesize,
  frame=lines,  
  framesep=2em, 
  rulecolor=\color{Gray},
  label=\fbox{\color{Black}proB encoding},
  labelposition=topline,
}

\author{Matthias van der Hallen\dag
\thanks{Matthias van der Hallen is supported by a Ph.D. fellowship from the Research Foundation - Flanders (FWO - Vlaanderen).\vspace{-3em}}
, Sergey Paramonov\dag, Michael Leuschel\ddag, Gerda Janssens\dag}
\title{Knowledge Representation Analysis of Graph Mining}
\institute{\dag KU Leuven, \ddag Heinrich-Heine-Universit\"at D\"usseldorf}

\begin{document}
\maketitle
\begin{abstract}
Many problems, especially those with a composite structure, can naturally be expressed in higher order logic. 
From a KR perspective modeling these problems in an intuitive way is a challenging task.
In this paper we study the graph mining problem as an example of a higher order problem. 
In short, this problem asks us to find a graph that frequently occurs as a subgraph among a set of example graphs.
We start from the problem's mathematical definition 
to solve it
in three state-of-the-art specification systems.
For IDP and ASP, which have no
native support for higher order logic, we propose the use of encoding
techniques such as the disjoint union technique and the saturation technique.
ProB
benefits
from the higher order support for sets.
We compare the performance of the three approaches to get an idea of
the overhead of the higher order support.

We propose higher-order language extensions for IDP-like
specification languages
and  discuss what kind of solver support is needed.
Native higher order shifts the burden of rewriting specifications using encoding techniques from the user to the solver itself.


\end{abstract}

\vspace{-1em}
\section{Introduction}
Many real world problems exhibit a composite structure consisting of multiple smaller problems which can be combined in many different configurations.
These types of problems lend themselves for a declarative approach as knowledge representation offers a transparent, natural and extendable model satisfying `The Principle of Elaboration Tolerance'~\citep{elaboration_tolerance}: 
declarative specifications are 
easily
adapted to new requirements or changed circumstances, e.g. variations in which subproblems are used, and in the way they are combined.
Conversely, the smaller problems in these composite structures are often already NP or coNP complete.
Combining these already complex problems often raises the computational complexity of the composite problem, up to a level where it cannot be expressed using first order logic.
These problems become higher order logic problems: We study the \emph{Graph Mining} problem as an example featuring such a raise in complexity.

Specification languages with support for higher order logic exist, with different levels of support.
On the one hand, meta-programming, as known from Logic Programming~\citep{abramson1989meta}, has inspired the introduction of higher-order atoms in DLVHex~\citep{conf/ijcai/EiterIST05}  and the higher-order syntax in HiLog~\citep{chen1993hilog}.
As in Prolog, predicate symbols can be either constants (first order case) or variables (second order case).
In the case of predicate variable symbols, these variables range over predicate names, and not the predicate space itself, essentially combining second order syntax with first order semantics. 
This cannot model the graph mining problem.

On the other hand, formal specification languages such as Z \citep{Bowen:Z}, B \citep{Abrial:BBook}, Event-B~\citep{Abrial10} and TLA~\citep{books/aw/Lamport2002}
 extend predicate logic with set theory and offer higher order datastructures. ProB~\citep{journals/sttt/LeuschelB08} is a constraint solver, animator and model checker for such languages, implemented in SICStus Prolog.
We can express the graph mining problem in ProB directly using higher order logic, 
but in general such systems miss the flexibility to perform multiple different inferences such as model expansion and optimization without modifying the specification.
Furthermore, ProB requires an encoding to express inductive definitions,
and as it is built on CP techniques and finite domain solvers, it does not benefit from the recent revolutions in solving techniques such as CDCL.

Therefore, we also look at specification languages that do not allow higher order syntax.
Examples of such languages are the IDP~\citep{WarrenBook/DeCatBBD16} and the ASP~\citep{conf/rweb/EiterIK09} language.
For these languages, several techniques exist that allow the user to simulate higher order logic to model problems such as graph mining, potentially offering better performance than systems that allow higher order logic directly.



Graph mining is a specific kind of \emph{frequent pattern mining}, 
the task of enumerating patterns which occur frequently in a dataset.
A first class of \emph{pattern mining} is \emph{unstructured mining}, such as \emph{itemset mining}, where the pattern is a set of items without any additional structural relation between the different items. 
This problem is of propositional nature:
\citet{tias_original} modeled it using CP techniques, while \citet{asp_itemset} used ASP.
Recently, focus has shifted from unstructured towards structured mining, such as graph or sequence mining \citet{cp_sequence_mining,asp_sequence}.
Here, the items being mined exhibit additional structure, for example the edge relation in the case of graph mining.
This introduces the\NP-coplete problem of graph homomorphism~\citep{Lev73},
and its many variations, which in imperative languages lead to many different algorithms~\citep{gspan,theta_subsumption}.
A declarative approach can express these variations with only minimal changes.

In our case study of the graph mining problem, we start with from the mathematical model of graph mining, which is inherently higher order, and identify the following contributions:
\begin{compactitem}
\item We identify the higher order aspects of the graph mining problem and show how the problem can be modeled in IDP, ASP and ProB, proposing concrete modeling techniques.
We also identify a set of desirable properties for a declarative encoding of the graph mining problem.
\item 
We propose a higher order encoding that closely follows the mathematical model of graph mining, and satisfies all desirable properties of a declarative graph mining model.
We indicate how additional solver support can exploit the additional structure in this encoding to work more efficiently.
\end{compactitem}
The paper is structured as follows: Section \ref{sec:formalization} introduces graph mining formally, Section \ref{sec:modeling} discusses the how to model the problem in IDP, ASP and ProB, identifying a set of desirable properties.
Then, Section~\ref{sec:performance} discusses the performance of these systems.
Section \ref{sec:extension} discusses a faithful encoding of the graph mining problem in an KR language enriched with HO, and its possible solver implementation. Section \ref{sec:conclusion} draws conclusions and outlines possible future research directions.

\section{Formalization of the graph mining problem}\label{sec:formalization}

\subsection{Patterns}
We start with a comprehensive formal definition of the graph mining problem.

\begin{definition}
A labeled graph \graph{G} is a triple $\triple{V,E,l}$ where $V$ is the finite set of vertices or nodes, $E$ is a binary predicate on $V$ that represents the set of (directed) edges and $l$ is a unary function from $V$ to a set of labels.
\end{definition}

\begin{definition}
A graph $\graph{G} = \triple{V,E,l}$ is \emph{connected} iff for each pair of vertices $v$ and $v'$ in $V$, there exists an edge $\pair{v,v'} \in E$ or there exists a sequence $v, v_{1} \ldots v_{n}, v'$ such that there exist edges $\pair{v,v_{1}}$, $\pair{v_{i},v_{i+1}}$ and $\pair{v_{n},v'} \in E$, where $1 \leq i \leq n-1$.
\end{definition}

\begin{definition}
A graph homomorphism $f$ from a labeled graph $\graph{G} = (V,E,l)$ to a labeled graph $\graph{G}' = (V',E',l')$ is an injective mapping $f$ : $V \rightarrow V'$ from vertices of $\graph{G}$ to vertices of $\graph{G'}$ such that:
\begin{compactitem}
\item $\forall v \in V : l(v) = l(f(v))$ (the mapping respects labelings), and
\item $\forall u,v \in V, \pair{u,v} \in E \implies \pair{f(u),f(v)} \in E'$ (the mapping preserves edges).
\end{compactitem}
If a graph homomorphism from graph $\graph{G}$ to $\graph{G'}$ exists we say $\graph{G}$ is \emph{homomorphic} with $\graph{G'}$.
\end{definition}

\begin{definition}
\label{def:GM1}
Given a pair $\triple{\graphset{E}_{+},\graphset{E}_{-}}$ consisting of a set of \emph{positive} and \emph{negative} examples of \emph{labeled graphs} respectively, 
and a graph $\graph{T}$ called the template,
\emph{Graph mining} is the problem of finding a pattern $\graph{P}$ which is
\begin{compactitem}
\item a \emph{connected labeled subgraph of $\graph{T}$},
\item \emph{homomorphic} with at least $N_{+}$ positive examples $\graph{E}_{+} \in \graphset{E}_{+}$, while being homomorphic with at most $N_{-}$ negative examples $\graph{E}_{-} \in \graphset{E}_{-}$. 
\end{compactitem}
\end{definition}
We call these homomorphisms the positive (negative) homomorphisms, and the restriction on their number the positive (negative) homomorphic property, respectively.

\tikzstyle{vertex} = [circle, fill=black, radius=1pt, inner sep=0pt, minimum size=3pt]
\begin{figure}[h]
\vspace{-1em}
  \centering
  \begin{subfigure}[b]{0.24\textwidth}
    \centering
    \begin{tikzpicture}[scale=.25]
      \node[vertex] (a) at (1,1) {};
      \node[vertex] (b) at (2,1) {};
      \node[vertex] (c) at (2.7,2) {};
      \node[vertex] (d) at (2,3) {};
      \node[vertex] (e) at (1,3) {};
      \node[vertex] (f) at (0.3,2) {};
      \draw (1,1) -- (2,1) -- (2.7,2) -- (2,3) -- (1,3) -- (0.3,2) -- (1,1);
      \draw (1,1) -- (2,3);
      \draw (2.7,2) -- (0.3,2);
    \end{tikzpicture}
    \caption{Positive Example\label{fig:pos}}
  \end{subfigure}
  ~
  \begin{subfigure}[b]{0.24\textwidth}
    \centering
    \begin{tikzpicture}[scale=.25]
      \node[vertex] (a) at (0,0) {};
      \node[vertex] (b) at (1,1) {};
      \node[vertex] (c) at (2,0) {};
      \node[vertex] (d) at (3,1) {};
      \coordinate (1) at (0,0);
      \coordinate (2) at (1,1);
      \coordinate (3) at (2,0);
      \coordinate (4) at (3,1);
      \draw (1) -- (2) -- (3) -- (4);
    \end{tikzpicture}
    \caption{Negative Example\label{fig:neg}}
  \end{subfigure}
  \begin{subfigure}[b]{0.24\textwidth}
    \centering
    \begin{tikzpicture}[scale=.25]
      \node[vertex] (a) at (1,1) {};
      \node[vertex] (b) at (2,1) {};
      \node[vertex] (c) at (2.7,2) {};
      \node[vertex] (d) at (2,3) {};
      \node[vertex] (e) at (1,3) {};
      \node[vertex] (f) at (0.3,2) {};
      \node[vertex] (g) at (3,3) {};
      \node[vertex] (h) at (3.7,3.7) {};
      \draw (1,1) -- (2,1) -- (2.7,2) -- (2,3) -- (1,3) -- (0.3,2) -- (1,1);
      \draw (1,3) -- (2,1);
      \draw (2,3) -- (3,3) -- (3.7,3.7);
    \end{tikzpicture}
    \caption{Template \label{fig:templ}}
  \end{subfigure}
  \begin{subfigure}[b]{0.24\textwidth}
    {
      \setcounter{subfigure}{0}
      \renewcommand\thesubfigure{\Roman{subfigure}}
      \centering
      \begin{subfigure}[b]{0.24\textwidth}
        \begin{tikzpicture}[scale=.25]
          \node[vertex] (a) at (1,1) {};
          \node[vertex] (b) at (2,1) {};
          \node[vertex] (c) at (2.7,2) {};
          \node[vertex] (d) at (2,3) {};
          \node[vertex] (e) at (1,3) {};
          \node[vertex] (f) at (0.3,2) {};
          
          \draw (1,1) -- (2,1) -- (2.7,2) -- (2,3) -- (1,3) -- (0.3,2) -- (1,1);
          \draw (2,1) -- (1,3);
        \end{tikzpicture}
        \caption{\label{fig:correctcandidate}}
      \end{subfigure}
      ~
      \begin{subfigure}[b]{0.24\textwidth}
        \begin{tikzpicture}[scale=.25]
          \node[vertex] (a) at (1,2) {};
          \node[vertex] (b) at (2,3) {};
          \node[vertex] (c) at (3,3) {};
          \draw (1,2) -- (2,3) -- (3,3);
        \end{tikzpicture}
        \caption{\label{fig:incorrectcandidate}}
      \end{subfigure}
    }
    \setcounter{subfigure}{3}
    \caption{Pattern Candidates\label{fig:candidates}}
  \end{subfigure}
  \caption{A graph mining instance ($N_{+}=1, N_{-}=0$) with pattern candidates.\label{fig:ex1}}
\end{figure}


Take, for example, the problem set shown in \textbf{Fig.}~\ref{fig:ex1}. We assume all nodes have the same label, and that all edges are bidirectional.
The template graph guides the search.
There is one positive example (\textbf{Fig.}~\ref{fig:pos}), and one negative example (\textbf{Fig.}~\ref{fig:neg}). 
\textbf{Fig.} \ref{fig:templ} shows the template graph.
\textbf{Fig.} \ref{fig:candidates} shows a valid and an invalid pattern. 
They are both connected subgraphs of the template.
Requiring at least one homomorphism with a positive example, and allowing no homomorphisms with negative examples (i.e. problem parameters $N_{+}=1$ and $N_{-}=0$), \textbf{Fig.} \ref{fig:correctcandidate} represents a valid pattern.
It is clear that there exists a mapping from each node from the valid pattern to a node of the positive example, while no such mapping exists for the negative example.
Looking at \textbf{Fig.} \ref{fig:incorrectcandidate}, this graph is clearly homomorphic with both the positive as well as the negative example. Therefore, it is not a pattern.

\subsection{Canonical patterns}
To extend on the graph mining task described above, we can look for multiple patterns, instead of just one.
In this case, one can impose restrictions on the different patterns that are found.
For example, it stands to reason that one wants only \emph{canonical} solutions, meaning that no two patterns found are \emph{isomorphic}.

\begin{definition}
\label{def:isomorphism}
A graph isomorphism $f$ between two labeled graphs $\graph{G} = \triple{V,E,l}$ and $\graph{G'} = \triple{V',E',l'}$ is a \emph{one-to-one} mapping $V \rightarrow V'$ 
such that $f$ represents a homomorphism from $\graph{G}$ to $\graph{G'}$,
and its inverse $f^{-1}$ represents a homomorphism from $\graph{G'}$ to $\graph{G}$.
If there exist graph isomorphisms between $\graph{G}$ and $\graph{G'}$ we say $\graph{G}$ and $\graph{G'}$ are \emph{isomorphic}.
\end{definition}

\vspace{-2em}

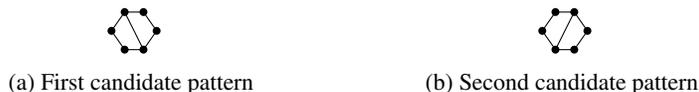
\begin{figure}[h]
  \centering
  \begin{subfigure}[b]{0.45\textwidth}
    \centering
    \begin{tikzpicture}[scale=.25]
      \node[vertex] (a) at (1,1) {};
      \node[vertex] (b) at (2,1) {};
      \node[vertex] (c) at (2.7,2) {};
      \node[vertex] (d) at (2,3) {};
      \node[vertex] (e) at (1,3) {};
      \node[vertex] (f) at (0.3,2) {};   
      
      \draw (1,1) -- (2,1) -- (2.7,2) -- (2,3) -- (1,3) -- (0.3,2) -- (1,1);
      \draw (2,1) -- (1,3);
    \end{tikzpicture}
    \caption{First candidate pattern\label{fig:iso1}}
  \end{subfigure}
  ~
  \begin{subfigure}[b]{0.45\textwidth}
    \centering
    \begin{tikzpicture}[scale=.25]
      \node[vertex] (a) at (1,1) {};
      \node[vertex] (b) at (2,1) {};
      \node[vertex] (c) at (2.7,2) {};
      \node[vertex] (d) at (2,3) {};
      \node[vertex] (e) at (1,3) {};
      \node[vertex] (f) at (0.3,2) {};
 
      \draw (1,1) -- (2,1) -- (2.7,2) -- (2,3) -- (1,3) -- (0.3,2) -- (1,1);
      \draw (1,1) -- (2,3);
    \end{tikzpicture}
    \caption{Second candidate pattern\label{fig:iso2}}
  \end{subfigure}
  \caption{Possible patterns\label{fig:isomorphism}}
\end{figure}

\vspace{-1em}
Given the graph mining problem as specified in \textbf{Fig.} ~\ref{fig:ex1}, we have already established that \textbf{Fig.}~\ref{fig:iso1} is a valid pattern.
When we try to mine a second pattern, we might suggest a pattern as shown in \textbf{Fig.}~\ref{fig:iso2}.
A quick check, however, will show that there is a one-to-one mapping $f$ such that both $f$ as well as its inverse $f^{-1}$ preserve edges.
As a result, both candidate patterns are isomorphic, and thus only one should be accepted as a valid pattern.

\subsection{Rewording}
We want to study how this formal mathematical definition can be expressed in the logics underlying the IDP~\citep{WarrenBook/DeCatBBD16} and the ProB~\citep{LeuschelSchneider_ABZ14} system.
First, we will reword the earlier \textbf{Def.}~\ref{def:GM1} into an equivalent formal definition that uses logical sentences and language constructs available in general logics.
In doing this, it becomes evident that the graph mining problem has fundamental underlying characteristics that result in a higher order definition and specification.


The vertices in the graph mining problem have no distinctive property, and can be reused between different example graphs and patterns.
Therefore, we will assume one shared, sufficiently large set of vertices $V$ and represent example graphs over these vertices $V$ directly as triples $\triple{Edge, Label, Class}$, consisting of an (binary) edge relation on $V$ and a labeling function over $V$, as well as a classification (positive/negative).

\begin{definition} \textbf{Graph Mining (redefined)}
\label{def:gm2}
Given a sufficiently large set of vertices $V$, a set $\graphset{G}$ of graphs over this vertex set $V$, represented by $\triple{E, l, c}$ triples
where $E$ and $l$ represent the edge relation and labeling function over $V$ respectively, and a template graph $\graph{T}$,
we look for a 
graph \graph{P} represented by tuple $\triple{E_{\graph{P}}, l_{\graph{P}}}$ such that:
\vspace{0.5em}
\begin{compactitem}
\item $\graph{P}$ is a \emph{connected} subgraph of $\graph{T}$,
\item $\#\Big\lbrace \triple{E,l,pos} \in \graphset{G} \; | \; \exists f : \text{f is a homomorphism from \graph{P} to \triple{E,l,pos}} \Big\rbrace \geq N_{+}$,

\item $\#\Big\lbrace \triple{E,l,neg} \in \graphset{G} \; | \; \exists f : \text{f is a homomorphism from \graph{P} to \triple{E,l,neg}} \Big\rbrace \leq N_{-}$.
\end{compactitem}
\end{definition}

\begin{definition} \textbf{Canonical Patterns}
A set of \emph{canonical patterns} is a set $\graphset{P}$ of graphs $\graph{P}_{1},...,\graph{P}_{n}$, such that for each pair of different elements (of \graphset{P}) $\graph{P}_{i}, \graph{P}_{j}$ holds that there does not exist an isomorphism between $\graph{P}_{i}$ and $\graph{P}_{j}$.
\end{definition}

Graphs are the main concept in the graph mining problem, and, when represented using  triples $\triple{E,l,c}$, graphs take the form of \emph{higher order objects}.
A set of graphs
is equivalent to a set of triples.
The most straightforward representation of such a set would 
be a ternary predicate.
As the domains of this predicate range over predicates and functions, it is a higher order predicate.

It is very natural to consider and represent each graph as a \emph{coherent} ensemble of its own components: all characteristics (edges, labeling \ldots) of a graph are represented by separate entities or concepts, which are grouped together for each graph $\graph{G}$ in the triple that describes it. We refer to this as the \emph{local coherence} of the graph representation.
Not only is this a very natural representation, this representation also makes it very explicit that all example graphs are \emph{independent}, and that the searches for homomorphisms between a pattern and example graphs are independent as well.
This motivates us to reason about graphs as locally coherent objects in our logical models as well.
However, the higher order representations needed to reason about graphs and sets of graphs as \emph{coherent} objects in our models are not yet fully supported by the logics of IDP and ASP.
In the following section discusses how to solve this using several modeling techniques.




\section{Modeling}\label{sec:modeling}
In this section, we show how state-of-the-art KR systems without support for higher order logic, such as IDP and ASP, can model the graph mining problem and its higher order features using encoding techniques.
We identify the desirable properties that from a KR perspective should hold for a good modeling of the graph mining problem and we evaluate how a modeling in ProB, as a KR language with support for higher order sets, satisfies these properties.

\subsection{IDP}
\subsubsection{Existential Second Order}
The IDP language can express problems that consist of a set of symbols, called the vocabulary $V$, and a theory, called $T$, that uses symbols from this vocabulary.
The symbols in the vocabulary can be propositions, but they can also represent predicates and functions.
These last two types of symbols make the vocabulary, in general, a \emph{second order} object: it is an object that itself \emph{contains} not only propositional symbols, but also first order symbols.
For example, vocabulary $V$ in \textbf{Listing} \ref{lst:vocabularyExample} is a second order vocabulary as it contains the first order symbol \lstinline{Edge/2}.

The theory $T$ is restricted to a \emph{first order} theory, extended with types, arithmetic, aggregates, and inductive definitions.
An example of such a theory is given in \textbf{Listing} \ref{lst:vocabularyExample}.
It contains an inductive definition for \lstinline{Path/2}, and one constraint.

Our inference of choice in the graph mining problem is model expansion; we search for an interpretation $I$ of symbols in the vocabulary $V$, called a \emph{model}, such that this interpretation $I$ satisfies the theory $T$.
This corresponds to the implicit \emph{existential quantification} of all symbols in the vocabulary, both the propositional as well as the first order symbols.
In the example of \textbf{Listing}~\ref{lst:vocabularyExample}, we expand the given interpretation $S$ to the model $Result$ with 3 edges: One from the first node to itself, one from the first node to the second, and one from the second to the third.
Path contains all corresponding paths between these three nodes.

In conclusion, we say the IDP language can express model expansion for \emph{Existential Second Order} problems.
This level of expressiveness is not sufficient for general graph mining problems.

\begin{lstlisting}[mathescape,style=model,caption={IDP example using inductive definitions}, label=lst:vocabularyExample]
vocabulary V{
    type Node, 
    Edge(Node, Node), Path(Node, Node)
}
theory T : V {
    $\forall$n[Node] : $\exists$n2[Node] : Edge(n,n2) $\lor$ Edge(n2,n).
    {
        Path(x,y) $\leftarrow$ Edge(x,y).
        Path(x,y) $\leftarrow \exists$z[Node] : Path(x,z) $\land$ Path(z,y).
        Path(x,y) $\leftarrow$ Path(y,z).
    }
}
structure S : V{ Node = {1;2;3} }
structure Result : V{
    Node = {1; 2; 3}, Edge = {1,1; 1,2; 2,3}
    Path = {1,1; 1,2; 1,3; 2,1; 2,2; 2,3; 3,1; 3,2; 3,3 }
}
\end{lstlisting}

\paragraph{Issue 1}
First, we must represent the set of example graphs, as specified in \textbf{Def.}~\ref{def:gm2}. 
This definition uses a higher order predicate \lstinline{GraphInst/3} (See \textbf{Listing}~\ref{lst:HOPred}) with the edge predicate as first argument and the labeling function as second argument. For the first graph, \lstinline|{1,2; 2,1}| and \lstinline[mathescape]|{1$\mapsto$ a; 2$\mapsto$ b}| respectively.
It represents a single graph as a tuple of predicates and functions, which is a highly locally coherent representation, preserving the independence of graph characteristics.
However, as we are restricted to \emph{Existential} Second Order, we cannot express this higher order predicate in IDP.

One possible solution is to replicate for each graph the different characteristic predicates and functions, as shown
in \textbf{Listing}~\ref{lst:multiglobal}.
In this encoding, every graph has its own edge predicate and label function. 
Because there is now no relation between the different edge predicates and label functions, it is necessary to formulate our theory in terms of these different predicates and functions.
Encoding a property such as ``In every graph, all nodes have at least two outgoing edges'' must be stated for each of the edge predicates explicitly:
\begin{center}
\begin{tabular}{c}
\begin{lstlisting}[mathescape]
$\forall$ n[Node] : $\exists$ n1,n2[Node] : E1(n, n1) $\land$ E1(n,n2) $\land$ n1 $\neq$ n2.
$\forall$ n[Node] : $\exists$ n1,n2[Node] : E2(n, n1) $\land$ E2(n,n2) $\land$ n1 $\neq$ n2.
\end{lstlisting}
\end{tabular}
\end{center}

It is clear that this solution is undesirable due to the way it scales and the theory modifications needed with growing problem instances.
It retains the local coherence and independence of graph characteristics when it comes to data representation, but prohibits the abstraction (generalization) of knowledge in the theory.
\vspace{-1em}
\begin{center}
\begin{minipage}{0.80\linewidth}
  \begin{lstlisting}[style=small,mathescape,caption=Higher order predicate modeling the set $\graphset{G}$ of \textbf{Def.}~\ref{def:gm2}.,label=lst:HOPred]
GraphInst({1,2; 2,1},{1$\mapsto$a; 2$\mapsto$b},pos).
GraphInst({1,3; 2,1},{1$\mapsto$c; 2$\mapsto$b; 3$\mapsto$a},neg).
\end{lstlisting}
\end{minipage}
\end{center}
\vspace{-1em}

\begin{minipage}[t]{0.45\linewidth}
\begin{lstlisting}[style=small,mathescape,caption=Multiple individual global relations,label=lst:multiglobal]
E1(1,2). lb1(1)=a.
E1(2,1). lb1(2)=b.
E2(1,3). lb2(1)=c.
E2(2,1). lb2(2)=b.
         lb2(3)=a.
\end{lstlisting}
\end{minipage}
\begin{minipage}[t]{0.1\linewidth}
 ~
\end{minipage}
\begin{minipage}[t]{0.45\linewidth}
\begin{lstlisting}[style=small,mathescape,caption=Disjoint union using indexed global relations,label=lst:indexedglobal]
E(g1,1,2). lb(g1,1)=a.
E(g1,2,1). lb(g1,2)=b.
E(g2,1,3). lb(g2,1)=c.
E(g2,2,1). lb(g2,2)=b.
           lb(g2,3)=a.
\end{lstlisting}
\end{minipage}

A more workable solution is to represent each characteristic property, such as the edge relation, by a single global relation for all graphs, as shown in \textbf{Listing}~\ref{lst:indexedglobal}.
This relation behaves the way it should for a specific graph instance based on an additional argument serving as an identifier for the graph of interest.
This global edge relation now corresponds to the \emph{disjoint} or \emph{tagged union} of the graphs' edge relations, where the tags are drawn from a set $G$ consisting of graph identifiers.
It is clear that this representation forces us to give up the local coherence of graph characteristics that was present in \textbf{Def.}~\ref{def:gm2}.
%
However, generalizing over the different graphs, we can now encode the property stated above as:
\begin{center}
\begin{minipage}{1.02\linewidth}
\vspace{-2em}
\begin{lstlisting}[style=small,mathescape]
$\forall$ gid[GraphId] : $\forall$ n[Node] : $\exists$ n1,n2[Node] : E(gid, n, n1) $\land$ E(gid, n,n2) $\land$ n1 $\neq$ n2.
\end{lstlisting}
\end{minipage}
\vspace{-3em}
\end{center}

\paragraph{Issue 2}
The homomorphic property can be expressed using a count aggregate, as shown in \textbf{Listing}~\ref{lst:QuantifyOutsideVocabulary}. 
First we quantify over all example graphs $g$, or per \textit{Issue 1}, their identifiers, and subsequently express that there must exist a function $f$ that represents a homomorphism from our pattern graph $\graph{P}$ to $g$.
\begin{center}
\begin{minipage}{0.75\linewidth}
\begin{lstlisting}[style=small,mathescape, caption=Quantifying over functions outside the vocabulary, label=lst:QuantifyOutsideVocabulary]
#{$g \mid g \in G \land \exists$ $f$ : $f$ is a homomorphism from $\graph{P}$ to $g$} $\geq$ N$_{+}$
\end{lstlisting}
\end{minipage}
\end{center}
However, IDP restricts us to Existential Second Order, which forbids us from quantifying over first order entities such as the function $f$ from \textbf{Listing}~\ref{lst:QuantifyOutsideVocabulary} outside of the vocabulary.
Thus, we are required to promote the homomorphic mapping functions to a global property in the vocabulary, even though we are only interested in the existence of a mapping, and not in the concrete instance of the mapping itself.
We prevent the same explosion of mapping functions as with the graph characteristics in \textit{Issue 1}, by reusing the disjoint union technique proposed above. 
Note that in this case, the disjoint union technique greatly resembles Skolemization. 
We introduce a general function \verb|f| that represents all homomorphisms, and make its dependency on a specific example graph explicit using an additional argument:
\verb|f(graphId, node):node|.
In Second Order Logic, this dependency would follow directly from the syntactic order of the quantifications.

\begin{center}
\begin{minipage}{0.66\linewidth}
\vspace{-1em}
\begin{lstlisting}[style=small,mathescape, caption=Globalized existential functions, label=lst:GlobalizeExistentialQuantifications]
#{$g \mid g \in G$ : $f(g)$ is a homomorphism from $\graph{P}$ to $g$} $\geq$ N$_{+}$
\end{lstlisting}
\vspace{-1em}
\end{minipage}
\end{center}

We can now use this \verb|f| anywhere we would use the regular homomorphic function for a specific graph by fixing the chosen example graph.
We denote by $f(g)$ the function $f$ partially applied on argument $g$.
Because the disjoint union technique introduces a single function $f$ which is the union of all these smaller functions, function $f$ becomes partial: it is not defined for tuples where the first the argument is an identifier for a graph $\graph{G}$ for which no homomorphic function exists.


\paragraph{Issue 3} The problem of deciding whether a homomorphism from one graph to another exists is \NP-complete.
As a result, deciding that no homomorphism from one graph to another exists, which forms the basis for the negative homomorphic property, is \coNP.
As an \NP\ (or $\Sigma^{p}_{1}$) solver, IDP cannot solve this problem directly.
The straightforward encoding of the negative homomorphic property reuses the result from \textit{Issue 2}:
\begin{center}
\begin{minipage}{0.65\linewidth}
\vspace{-1em}
\vspace{-1em}
\begin{lstlisting}[style=small,mathescape]
#{$g \mid g \in G$ : $f(g)$ is a homomorphism from $\graph{P}$ to $g$}$\leq$ N$_{-}$
\end{lstlisting}
\vspace{-1em}
\end{minipage}
\end{center}
But now, our solver must choose a single global function $f$ which satisfies the constraints.
It has no obligation to maximize the number of homomorphisms in $f$, only to satisfy the constraints.
Thus, even if there is a negative example $\graph{G}_{-}$ for which a homomorphism exists, the solver can choose $f$ such that $f$ does not represent a homomorphism for this graph $\graph{G}_{-}$.
As our constraints are satisfied, we are led to believe that our pattern candidate is a valid pattern.

\citep{conf/fsttcs/Immerman98} has shown that this is inherently linked to IDPs limit to Existential Second Order.
Indeed, in order to check that our pattern \graph{P} is homomorphic with no more than $N_{-}$ negative graphs, we have to check that there are enough negative graphs for which no homomorphism exists, for example using a count aggregate as in \textbf{Listing}~\ref{lst:universalquant}.
By asserting a property for all candidate homomorphic functions $f$ of a certain graph $g$, the negative homomorphic constraint leads to universal quantification over a function variable.

\vspace{-0.5em}
\begin{center}
\begin{minipage}{0.75\linewidth}
\begin{lstlisting}[style=small,mathescape, caption=Quantifying over functions outside the vocabulary, label=lst:universalquant]
#{$g \mid g \in G \land \forall$ $f$ : $f$ is $\mathbf{not}$ a homomorphism from $\graph{P}$ to $g$}
\end{lstlisting}
\end{minipage}
\end{center}
\vspace{-0.5em}

A way to solve a \coNP\ problem such as the negative homomorphism constraint using an \NP\ solver is by encoding the dual (i.e. negated) problem, and conclude that the problem is satisfied if no model exists for the dual problem.
This can be checked using an \NP\ solver.
However, this technique can only be implemented in IDP by writing two theories: 
\begin{compactitem}
\item one (positive) theory $\theory^{+}$ (see \ref{app:Code}), which expresses the positive homomorphic property and generates pattern candidates, and
\item one negative theory $\theory^{-}$, which expresses the (dual of) negative homomorphic property and rejects pattern candidates that do not satisfy this constraint.
\end{compactitem}
In IDP, one must provide procedural (lua) code that ties these two theories and their inferences together by allowing the communication of pattern candidates between these two theories.

It is not known whether the problem of graph isomorphism is polynomial time solvable,
however it is sure to be no more complex than NP.
Conversely, the isomorphism restriction when looking for multiple patterns is also no more complex than \coNP.
Therefore, we can use the same technique, giving rise to another theory $\theory^{iso}$.
Note that it is possible to combine the negative theories $\theory^{-}$ and $\theory^{iso}$ into a single negative theory.

\subsubsection{Inductive Definitions}
One of the main features of the IDP language is the fact that it extends first order logic with \emph{inductive definitions}. These definitions, evaluated under the well-founded semantics, allow the derivation of negative knowledge that otherwise would be underivable.
Take the path predicate defined in \textbf{Listing}~\ref{lst:vocabularyExample}.
Models of this theory contain the transitive closure \lstinline|Path|/2 of \lstinline|Edge|/2.
When the edge relation would be chosen such that two nodes $a$ and $b$ are part of two disconnected graphs, there is no model in which \lstinline|Path(a,b)| holds. Note that when the transitivity property is expressed as an FO constraint instead, there do exist models in which \lstinline|Path(a,b)| is true.


\subsubsection{Other inferences}
One of the advantages of IDP is its underlying \emph{Knowledge Base} paradigm~\citep{WarrenBook/DeCatBBD16}.
Essentially, this paradigm ensures that we can perform other inferences on the graph mining problem.
One of these inferences is, for example, optimization.
This would allow us to, e.g., minimize or maximize over the number of nodes in the pattern graph, or the number of nodes in the pattern with a certain label, with only minimal changes to the specification.

\reversemarginpar

\subsection{ASP}
In \textbf{ASP}, a language family closely related to IDP, one would mostly encounter the same issues when modeling the graph mining problem.
One of the main differences between ASP and IDP is the choice of semantics: ASP looks for the
answer set models, whereas IDP looks for well-founded models.
Leveraging the minimality property of answer sets, ASP can prevent the invalid models of the example discussed in \textit{Issue 3}.
The corresponding technique is called the \emph{saturation} technique~\citep{conf/rweb/EiterIK09} and can prevent the creation of two separate theories and writing of procedural code that IDP requires.

When using this technique, ASP detects negative example graphs for which the $f$ does not represent a homomorphism, and requires for these example graphs that $f$ must map every node of the pattern on every node of that example graph, dropping the injectivity constraint.
This way, $f$ becomes so large that it is impossible that it belongs to the minimal answer set unless there does not exist a homomorphism from the pattern to this (negative) example graph.
Consequently, the minimality property will cause the solver to look for an $f$ that represents a homomorphism for as many example graphs (including negatives) as possible.
The same technique can be applied to the isomorphism restriction and other possible $\Sigma_{2}^{p}$ constraints such as subset minimality.

While this technique successfully prevents the need of a procedural loop and the rewriting of the negative homomorphic property and the isomorphism restriction, it is clear that this technique is not derived from a natural KR translation of the Graph Mining definition.
Furthermore, as line~\ref{lstline:probspec} of \textbf{Listing}~\ref{lst:aspsaturation} (See \ref{app:Code}) shows, it is necessary to encode instance specific knowledge into the model.

\subsection{ProB} \label{subsection:prob}
The ProB System can handle mathematical specifications using higher order logic and set theory.
As a result, ProB specifications can cover the polynomial hierarchy \textbf{PH}~\citep{DBLP:books/daglib/0095988}.

\subsubsection{Higher Order Logic}
Because of ProB's Higher Order logic support, we can treat graphs as the inherent higher order objects with structure $\triple{E,l,c}$ that represents them.
This allows us to quantify over a graph and easily access all its characteristic predicates and functions.

ProB's higher order logic support also makes it possible to quantify over the functions $f$ that represent homorphisms locally: there is no need to declare the function $f$ globally, instead they are defined within the context of the set of homomorphic positive (negative) examples.
Here, the representation of these functions $f$ is direct, without graph identifier that corresponds to the disjoint union technique as proposed for IDP.
Instead, the graph \graph{G} for which a homomorphic function is sought, is brought in scope by the quantifier of the set expression.

Because these are now quantified locally, the solver will find a homorphism if one exists, regardless of whether we are expressing the positive or negative homomorphism property.
As a result, ProB can model the negative homomorphism property directly, without the need for a second theory and procedural tie-in code.

The same reasoning allows ProB to model the isomorphism restriction when looking for multiple patterns.
\subsubsection{Inductive definitions}
ProB does not support inductive definitions, but allows the expression these constraints using either the B transitive closure primitive or by expressing the completion of the definition.
However, these techniques tend to reduce the readability of the constraint, making it difficult for modelers to reason about the connectedness constraint and its derivatives.
Furthermore, these constraints incur a high performance loss.
Recently, efforts have been made to integrate Kodkod, which provides a high-level interface to SAT-solvers~\citep{DBLP:conf/tacas/TorlakJ07}, into ProB~\citep{DBLP:conf/fm/PlaggeL12}, which allows offloading these constraints to a SAT-solver that is capable of solving them fast.

\subsection{Comparative Summary} 

Using the graph mining problem as a case study, we derived a set of desirable properties that a good KR specification should satisfy.

\begin{enumerate}[itemsep=0mm]
\item Labeled graphs are the main concept in the mathematical definition of the graph mining problem. 
Here, labeled graphs are seen as a mathematical object consisting of an edge relation and a labeling function, and should be treated as higher order objects in the specification.
\item All example graphs are independent, so the search for a homomorphism between a pattern and a given example graph can be performed independently. 
In essence we want to allow \emph{local} second order quantification.  
\item The search for a homomorphism between pattern and example graph is always the same, regardless of the sign of the example graph (negative or positive). The only difference is the at most/at least constraint on the number of homomorphisms.
We want a specification that preserves the similarity of these constraints.
\item We want to be able to find multiple, non-isomorphic, patterns.
\item We want to express constraints such as connectedness of the different nodes in the pattern.
\item We want to perform multiple inferences on the problem, with only minimal changes to the model.
\item We prefer a single specification over multiple specifications. 
Although specifications are preferably modular to make it easier to reuse them, ideally the specification would be solved within a single solver call, requiring no procedural code to tie them together.
\end{enumerate}
\textbf{Table}~\ref{tbl:conclusion} provides an overview of how the three systems (IDP, ASP and ProB) support the desirable properties, either natively (\checkmark) or using one of the discussed techniques.
\begin{table}[h]
\begin{center}
\includegraphics[scale=0.75]{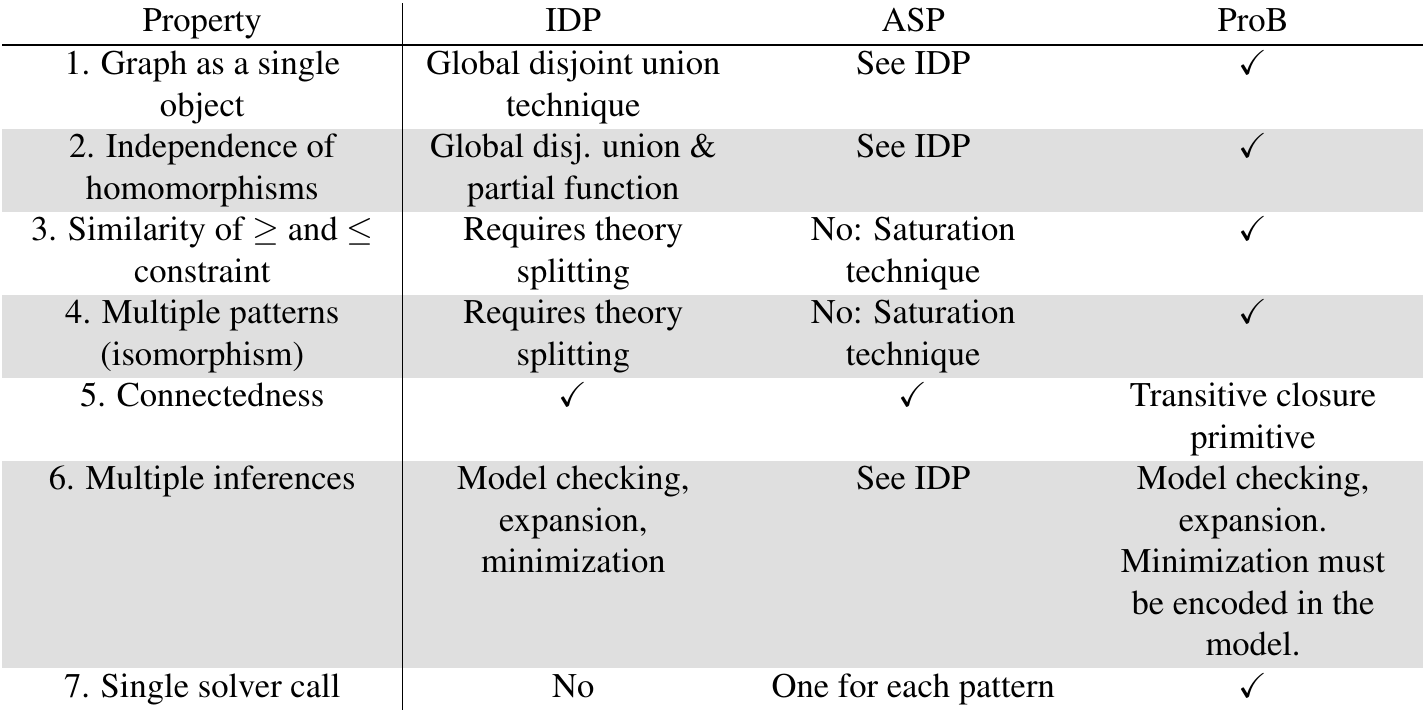}
\end{center}
\caption{Evaluation of the desirable properties in IDP / ASP / ProB\label{tbl:conclusion} for modeling the general graph mining problem based on its key components (matching, pattern enumeration, etc)}
\vspace{-3em}
\end{table}
\section{Performance}\label{sec:performance}
\vspace{-1em}
To compare the performance of higher order and first order systems, we compared the IDP system with the ProB system (which uses higher order specifications).
To this end, we used the positive examples of the Yoshida~\citep{yoshida_dataset} dataset, which is derived from biochemics, for graph mining.
First, we randomly picked an example to use as the template graph.
Next, we mined a pattern from this template, using the threshold value $N_{+} = 13$ (5\% of the size of the example set).
During the mining process, we tracked the time it takes to mine the $i=1..n$-th pattern. The results are averaged over ten runs.

The ProB model from Subsection \ref{subsection:prob} comes closest to the higher order formulation (as demonstrated in \textbf{Table} \ref{tbl:conclusion}), however, the solver support is not yet sufficient to efficiently execute the higher order graph mining model on larger datasets, i.e., currently we have not found an efficient way to mine patterns using a higher order B model. 
Consequently, from a KR point of view, we consider the higher order formulation of the graph mining problem as a challenge and goal for future solver techniques.
The key issue preventing an efficient higher order formulation lies in reifying the higher order existential quantifier inside the set comprehension. A possible future solution would be to provide a Prolog implementation for the homomorphism predicate (e.g., as a ProB external function).
For IDP, the results can be found in \textbf{Table} \ref{table:idp:averaged_results}.

\begin{figure}[thb]
\vspace{-2em}
\centering
\begin{subfigure}{.44\textwidth}
  \centering
\includegraphics[scale=0.14]{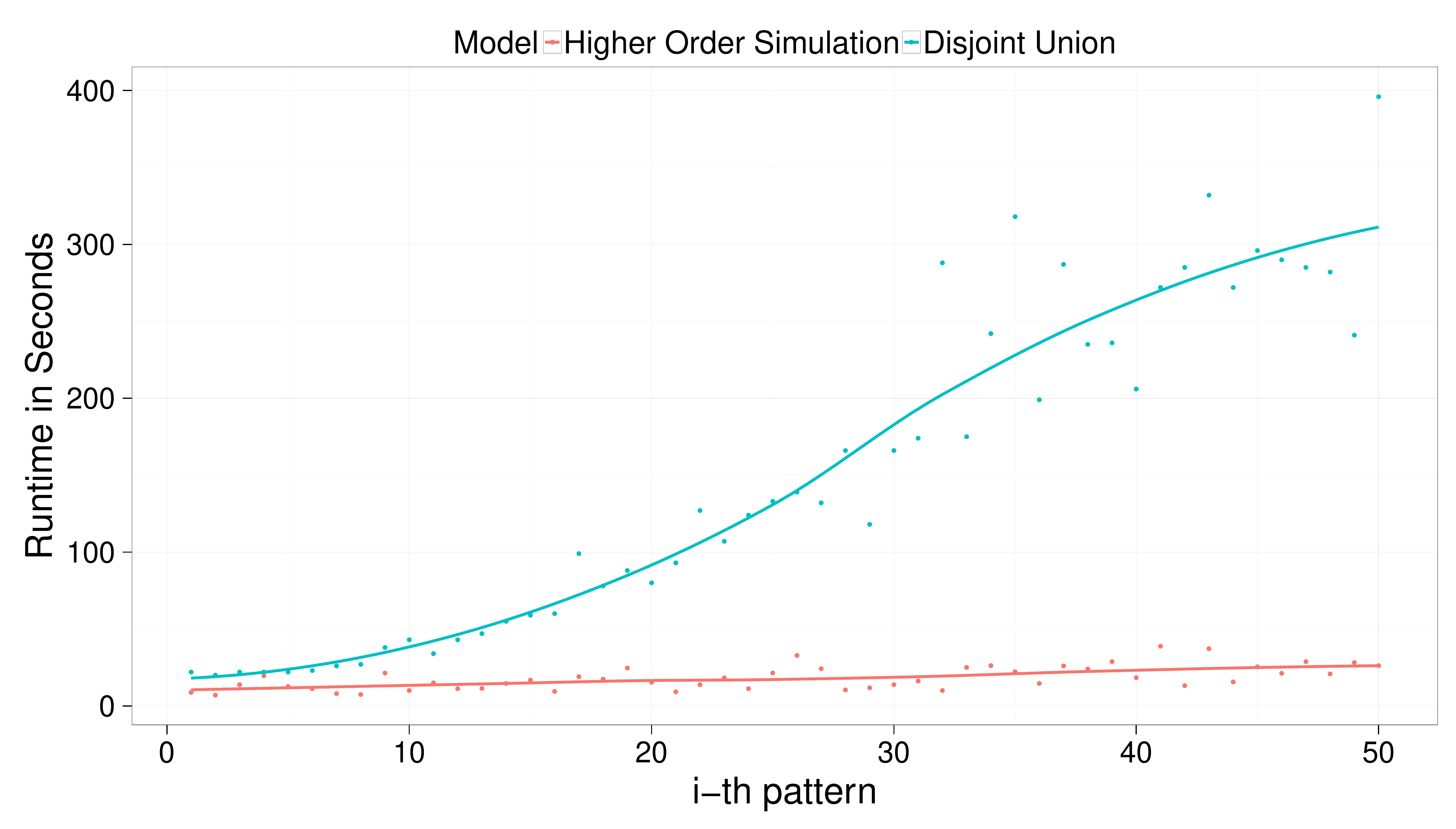}
\caption{\footnotesize{IDP: the disjoint model has a growing trend while the simulation stays flat. The gap is one order of magnitude. (\cite{ilp_graph_mining})}}
  \label{fig:decomposition_idp}
\end{subfigure}%
\hfill
\begin{subfigure}{0.46\textwidth}
  \centering
 \includegraphics[scale=0.14]{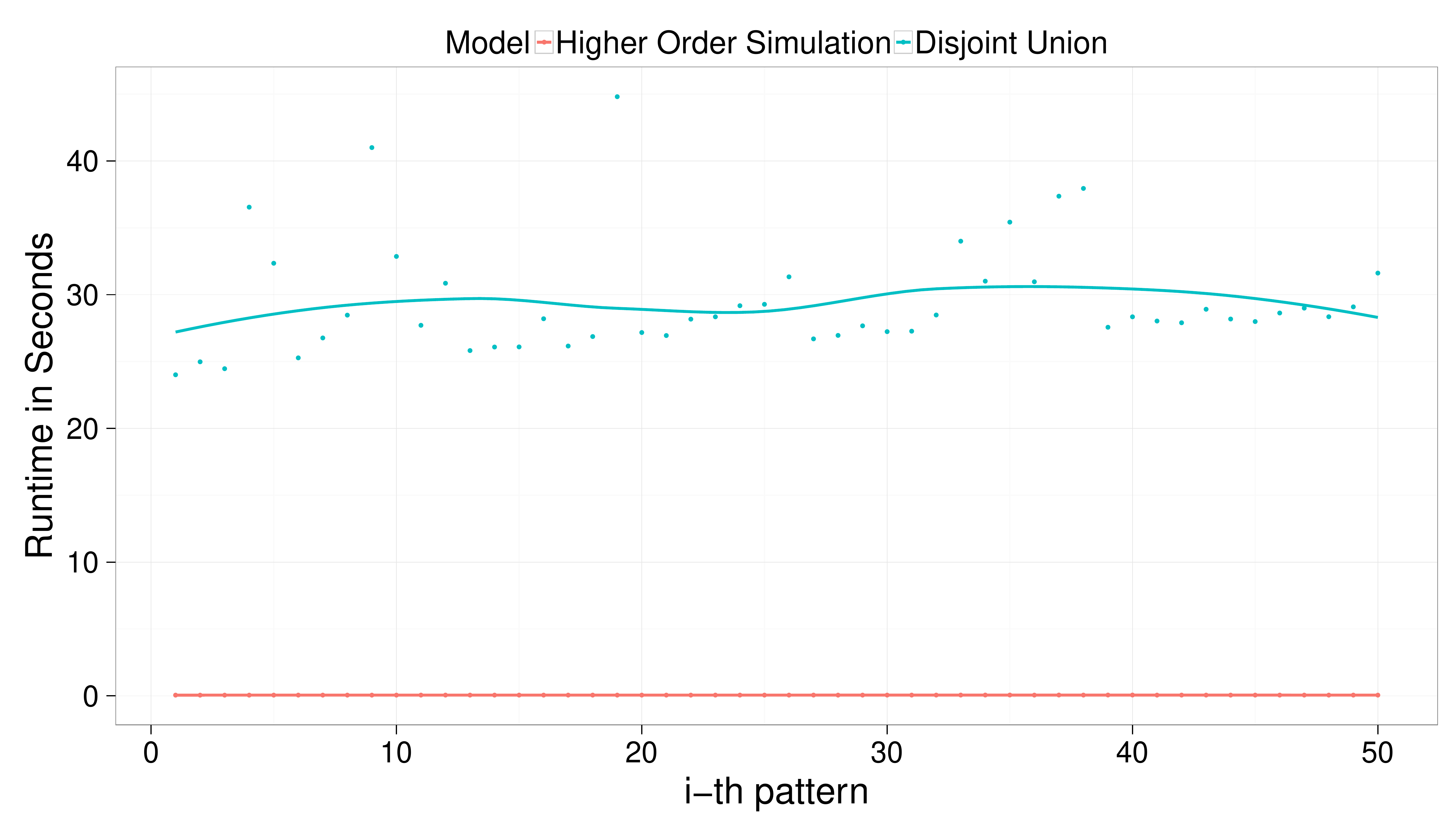}
 \caption{\footnotesize{ASP: the disjoint model exhibits fluctuation around 30s with a slow runtime growth, while the simulation stays flat. The gap is two orders of magnitude.}}
  \label{fig:decomposition_asp}
\end{subfigure}
\caption{\footnotesize{Frequent graph enumeration problem (5\% threshold) on Yoshida dataset for IDP (a) and ASP (b), comparing disjoint union (in blue) and higher order simulations (in red). Further details can be found in \ref{sec:hol_description}.}}
\label{fig:decomposition_fol}
\end{figure}

To analyze the effect of the disjoint union technique, we compared the performance of IDP and ASP on the Yoshida dataset using different encodings of the graph mining problem.
In \textbf{Fig.}~\ref{fig:decomposition_fol}, we see the performance of IDP (\textbf{Fig.}~\ref{fig:decomposition_idp}) and ASP (\textbf{Fig.}~\ref{fig:decomposition_asp}) on finding the $i$-th pattern.
Two different encodings are used: one that uses the disjoint union technique, and one that performs a new IDP/ASP call for every different example graph, and aggregates this data using procedural code (i.e. in a decomposed fashion).

It is clear from \textbf{Fig.}~\ref{fig:decomposition_fol} and the order(s) of magnitude difference between the decomposition and disjoint union technique that these systems can highly benefit from detecting the independence of these different subproblems and solving them separately.
We expect that expressing the problems in a higher order fashion will allow detection of this subproblem independence and allow for more performant and expressive systems.
\section{A faithful encoding}\label{sec:extension}


\vspace{-0.5em}
In \textbf{Listing}~\ref{lst:faithful}, we now propose a new encoding for a language combining higher order logic support with the readability of inductive definitions.
This encoding is more faithful to the problem with respect to the definition given in \textbf{Def.}~\ref{def:gm2}.

In the vocabulary, the second order type \lstinline{graph}, parametrized by two first order types \lstinline{node} and {\lstset{breakatwhitespace} \lstinline{label},} is declared as a tuple of a predicate \lstinline{vertex/1}, a predicate \lstinline{edge/2}, and a function \lstinline|label|.
Next, we declare the higher order predicates (\lstinline|homomorphism|, \lstinline|reachable|, \lstinline|isPattern|, \lstinline|canonical_pattern|, \lstinline|positive|, and \lstinline|negative|) and function (\lstinline|template|).

Within the theory, higher order predicates are defined using the concept of templates as described by \cite{DBLP:journals/tplp/DassevilleHJD15}. 
The higher order arguments are decomposed using matching (e.g. line~\ref{lstline:matching}) or using dot notation (e.g. line~\ref{lstline:dot}).
Quantification over second order objects uses annotated quantifiers ($\exists_{SO}$ and $\forall_{SO}$) and must be typed (any unary predicate represents a type), e.g. line~\ref{lstline:exists}.

\begin{lstlisting}[mathescape,style=model,caption={Faithful encoding for the general graph mining problem},label=lst:faithful, numbers=left]
Vocabulary V @\textbraceleft@
   type node, type label
   so-type graph(node, label) of (vertex(node), edge(node,node), label(node):label)
   homomorphism(graph, graph), reachable(node,node, graph)
   isPattern(graph), canonical_pattern(graph)
   positive(graph), negative(graph), template:graph 
@\textbraceright@ 
Theory T @\textbraceleft@ 
 @\textbraceleft@homomorphism((V1, Edge1, Label1), (V2, Edge2, Label2)) $\leftarrow$@\label{lstline:matching}@
     @\big(@$\exists_{SO}$ F [V1:V2] : ($\forall$ x, y : x $\neq$ y $\implies$ F(x) $\neq$ F(y)) $\wedge$@\label{lstline:exists}@
      ($\forall$ x, y : Edge1(x, y) $\implies$ Edge2(F(x), F(y))) $\wedge$
      ($\forall$ x : Label1(x) = Label2(F(x)))@\big)@.
 isomorph((V1, Edge1, Label1),(V2, Edge2, Label2)) $\leftarrow$
     @\big(@$\exists_{SO}$ F [V1:V2] : ($\forall$ y : y => $\exists$ x : F(x)=y) $\land$ 
      ($\forall$ x, y : x $\neq$ y $\implies$ F(x) $\neq$ F(y)) $\wedge$
      ($\forall$ x, y : Edge1(x, y) $\implies$ Edge2(F(x), F(y))) $\wedge$              
      ($\forall$ x, y : Edge2(x, y) $\implies$ $\exists$ fx, fy : Edge1(fx, fy) $\land$ x = F(fx) $\land$ y = F(fy))
      $\wedge$ ($\forall$ x : Label1(x) = Label2(F(x)))@\big)@.
 reachable(x, y, (Vertex, Edge, Label)) $\leftarrow$ Edge(x, y) $\lor$ Edge(y, x).
 reachable(x, y, (Vertex, Edge, Label)) $\leftarrow \exists$ z : reachable(x, z, (Vertex, Edge, Label)) $\wedge$ reachable(z, y, (Vertex, Edge, Label)).
 isPattern((Vertex, Edge, Label)) $\leftarrow$
     @\big(@($\forall x$: Vertex(x) $\implies$ template.vertex(x)) $\land$@\label{lstline:dot}@
      ($\forall x,y$: Vertex(x) $\wedge$ Vertex(y) $\wedge$ template.vertex(x) $\wedge$ template.vertex(y) $\wedge$ template.edge(x,y) $\implies$ Edge(x,y)) $\land$ 
      (@\#\textbraceleft@ Pos : positive(Pos) $\wedge$ homomorphism(P, Pos) @\textbraceright@ $\geq$ $N_{+}$) $\land$
      (@\#\textbraceleft@ Neg : negative(Neg) $\wedge$ homomorphism(P, Neg) @\textbraceright@ $\leq$ $N_{-}$) $\land$
      ($\forall$ x, y : reachable(x, y, P))@\big)@. @\textbraceright@
 $\forall$P : canonical_pattern(P) $\implies$ isPattern(P). 
 $\forall$P,P2 : canonical_pattern(P)$\wedge$canonical_pattern(P2)$\wedge$P$\neq$P2 $\implies$ $\neg$isomorph(P, P2).
@\textbraceright@ 
\end{lstlisting}

This encoding compactly specifies the graph mining problem, in a way that closely corresponds to its mathematical definition.
To allow inferences on this theory, extended solver support is necessary.
We now propose a way in which a solver can provide this additional support, and potentially even improve performance.
\vspace{-1em}
\subsubsection{Second order types}
The solver can represent objects of any \lstinline|so-type| using the disjoint union technique, declaring a new first order type $id$ containing identifiers for the higher objects, e.g. \lstinline|graphId|.
Using theory analysis, we determine whether the size of the second order type is bounded and if so, impose the same bound on the size of the type $id$.
If no such bound can be detected, we treat $id$ as an infinite type, relying on lazy grounding to create new $id$ objects when necessary and to subsequently instantiate the required rules for the new $id$ object.

Next, every occurrence of an object of type \lstinline|graph| is replaced by the correct identifier, and quantifications over this type are replaced by quantifications over the set of identifiers. Furthermore, every time a component of an object is accessed (e.g. \lstinline|Edge/2|) it is replaced by a global predicate
representing this component (i.e. \lstinline|Edge(gid, x, y)|).
\vspace{-1em}
\subsubsection{Second order quantifications $\exists_{SO}$/$\forall_{SO}$}
Second order quantifications such as $\exists_{SO}$ and $\forall_{SO}$ are supported using the concept of oracles as subsolvers.
First, all second order universal quantifications $\forall_{SO} X : \phi$ are rewritten to existential quantification $\neg \exists_{SO} X : \neg \phi$.
Suppose now that $\phi$ does not contain any further second order quantifications.
Then the above formula is an existential second order formula, which can be solved by a new instance of the \NP solver.
Recently, \cite{AAAIW1612603} have identified an interface by which any solver can be nested within another solver.
Because our \NP solver conforms to this interface, we can modify the \NP solver such that it calls a new instance of itself as an \emph{oracle} to evaluate the truth of these formulas.
The outer solver is called the \emph{top solver}, and the inner solver is called the \emph{subsolver} or \emph{oracle}.
As it is possible to nest these solvers arbitrarily deep, we can now solve a formula of the form $\exists_{SO} X : \phi$, regardless of whether $\phi$ contains any more second order quantifications.
Essentially, the \NP solver becomes a \QBF solver.

To set up a nested solver for a formula $\exists_{SO} X : \phi$, we must set up a vocabulary $V$ and a theory $T$ over $V$ for this solver. 
To this end, we first identify the variables $\Sigma$ used in $\phi$.
These variables $\Sigma$, together with the variable X from the quantification itself, are collected in the new vocabulary $V$. 
We call the free variables of $\phi$ the shared variables $\Sigma_s$.
We now use the formula $\phi$ as the theory $T$ for the subsolver.

Whenever the solver needs to evaluate the truth of a second order quantification, the solver simply calls this oracle on vocabulary $V$ and theory $T$, providing it with a set of assumptions consisting of the values that the top level solver assigns to the shared symbols $\Sigma_s$.
Depending on whether the subsolver succeeds or fails to find a model, we update the current interpretation of the top solver with the model or learn a new clause, as detailed by~\cite{AAAIW1612603}.
We expect this subsolver technique to allow detection of the independence of subproblems, thanks to the expressivity of higher order logic, and expect the performance of such a solver to close the gap with the performance of the decomposition technique detailed in Section~\ref{sec:performance}.
\section{Conclusion and future work}
\label{sec:conclusion}
\vspace{-1em}
In this paper we used graph mining as an example of a higher order problem and made a thorough analysis of the problem from the knowledge representation point of view. While techniques exist to express these higher order problems in first order logic, sometimes, explicitly specifying the additional structure HO exhibits allows systems to perform better. For example, in the case of graph mining, higher order logic preserves the local coherence of graphs, and the independence of homomorphisms for the different examples, a property that a higher order solver can exploit in order to raise efficiency. In its current state however, no technique combines the expressiveness of higher order logic with high performance solving techniques. 

Inspired by this case study, we propose higher-order language extensions for IDP and propose alternative ways to implement them in the solver. In particular, as shown in Section \ref{sec:performance}, the use of subsolvers seems promising and will be further explored together with the idea of Benders decomposition \citep{Benders}. The performance of the encodings in IDP or ASP can be considered as the ultimate target.

{\scriptsize
\bibliography{references}}
\bibliographystyle{abbrvnat}
\setcitestyle{authoryear,open={(},close={)}}

\newpage
\appendix
\section{Higher Order Logic Simulation Description}
\label{sec:hol_description}
Key dataset characteristics for the experiments, visualized in \textbf{Fig.} \ref{fig:decomposition_fol}, can be found in \textbf{Table} \ref{table:yoshida}.
\begin{table}[thb]
  \caption{Yoshida datasets parameters}
  \label{table:yoshida}
  \begin{tabular}{l c c c c c}
    Name & Number of Graphs & Avg Vertices & Avg Edges & Labels & Possible classes\\
    \hline
    Yoshida & 265 & 20 &  23 &  9 & 2
  \end{tabular}
\end{table}

The experimental setup for the results visualized in \textbf{Fig.} \ref{fig:decomposition_fol} is the following: in both disjoint union and higher order models we mined the patterns from smaller to larger in an iterative fashion. First, we set the pattern length, equal to the number of nodes, to two, then computed graph coverage for the pattern. Based on the coverage we add the pattern as frequent and then compute isomorphic patterns in the template. For each isomorphic graph in the template we add a no-good clause. Once all frequent patterns of the length $n$ are mined, i.e., the solver cannot find any other non-isomorphic patterns of the length $n$, we increase the pattern length to $n+1$, remove all no-goods and repeat the process.

The key difference between the disjoint union model and the higher order simulation model is in the coverage computation. In case of disjoint union model we make a single call to get a pattern such that it is frequent (i.e., matches at least the threshold amount of graphs) and in the higher order model we make a single call to get a non-isomorphic candidate graph and then a separate call per graph to find if it is covered or not. If we found that a pattern covers more than a threshold amount of graphs, we stop computing the coverage and add the pattern as frequent.

Both models in the described computations follow the general schema used in the specialized algorithms such as gSpan \citep{gspan}. We have also obtained similar runtime patterns on other standard graph datasets described in \citep{ilp_graph_mining}.

\section{IDP enumeration results}
In this section, we present the experimental results on the general graph mining IDP encoding using theory splitting (that allows incorporating positive, negative examples and other higher order checks in a uniform fashion). We have applied this encoding to Yoshida dataset on positive examples and used the isomorphism check as a negative theory. The results summarized in \textbf{Table} \ref{table:idp:averaged_results}. The results are consistent with the results in \textbf{Fig.} \ref{fig:decomposition_fol} of the more specialized encoding (that uses imperative code around the IDP/ASP calls) based on gSpan schema \citep{gspan}.

\begin{table}[h]
  \centering
  \setlength{\tabcolsep}{2pt}  
  \begin{tabular}{lccccccccccccccc}  Index  &   1 &   2 & 3 &   4 &   5    & 6   & 7   & 8   & 9   & 10  & 11  & 12  & 13 & 14 & 15\\ \hline 
 Runtime  & 148 & 164 & 173 & 199 &   285  & 364 & 401 & 445 & 490 & 533 & 548 & 585 & 591 & 687 & 802
  \end{tabular}
  \caption{Averaged runtimes in seconds for IDP general graph mining encoding with the theory splitting on the Yoshida dataset.} 
  \label{table:idp:averaged_results}
\end{table}

\section{Code}
\label{app:Code}
This appendix provides the relevant code for the IDP, ASP and ProB systems.
The full IDP code is available at\\ \url{https://dtai.cs.kuleuven.be/static/krr/files/experiments/aspocp16_IDP.zip} and at \\\url{https://github.com/SergeyParamonov/LGM},\\ while the ASP code is available at\\ \url{https://dtai.cs.kuleuven.be/static/krr/files/experiments/aspocp16_ASP.zip}\\ and the ProB code at \\\url{https://dtai.cs.kuleuven.be/static/krr/files/experiments/aspocp16_ProB.zip}. 
\begin{lstlisting}[caption=IDP positive constraint, style=model, label=lst:IDPPos]
vocabulary V{
  type node isa nat
  type graphid
  type label

  // Predicates determining the template graph.
  template_edge(node, node) 
  template_label(node):label

  // Predicates describing the positive example graphs
  example_edge(graphid, node, node)
  label(graphid, node):label
  threshold: int

  // Predicates describing the pattern graph
  inpattern(node) // True for the nodes which occur in the pattern
  partial f(graphid, node):node // Represents the homomorphisms with the example graphs
  homowith(graphid) // True for graphs for which f represents a correct homomorphism
  path(node, node) // path(a,b): True if there exists a path from a to b in the pattern
}

theory Positive:V_Pos{
   //The pattern is a connected subgraph of the template: From every node in the pattern, 
   //there exists a path to every other node in the pattern.
   !x,y[node] : x ~= y & inpattern(x) & inpattern(y) => path(x,y).
   {
      path(x,y) <- template_edge(x,y) & inpattern(x) & inpattern(y).
      path(x,y) <- ?z[node] : path(x,z) & path(z,y).
      path(x,y) <- path(y,x).
   }

   //existence of a homomorphic f from the pattern to example graph with graphid gid.
   !gid[graphid] : !x[node]   : homowith(gid) & inpattern(x) <=> ? y[node] : y=f(gid,x).
   !gid[graphid] : !x,y[node] : homowith(gid) & inpattern(x) & inpattern(y) & x~=y => f(gid, x) ~= f(gid,y).
   !gid[graphid] : !x,y[node] : homowith(gid) & inpattern(x) & inpattern(y) & template_edge(x,y) => edge(gid, f(gid,x). f(gid,y)).
   !gid[graphid] : !x[node] : homowith(gid) & inpattern(x) => template_label(x) = label(gid, f(gid,x)).

   // At least N homomorphisms must be found
   #{ gid [graphid] : homowith(graph) } >= threshold.
}
\end{lstlisting}

\lstset{basicstyle=\footnotesize\ttfamily,breaklines=true}
\begin{lstlisting}[caption=ASP positive matching, style=model]
0 { homowith(G) } 1 :- positive(G).

1 { f(G,X,V) : node(G,V) } 1 :- positive(G), inpattern(X).

:- used_f(G,X,V1), used_f(G,Y,V2), template_edge(X,Y), not edge(G,V1,V2), inpattern(X), inpattern(Y).
:- used_f(G,X,V),  t_label(X,L), not label(G,V,L), inpattern(X).

used_f(G,X,V) :- homo_with(G), f(G,X,V).
:- used_f(G,X,V), used_f(G,Y,V), X != Y.

positive_count(N) :- N = #count{G:homowith(G)}.

:- positive_count(N), N < 13.
\end{lstlisting}

\begin{lstlisting}[caption=ASP negative matching using saturation technique, label={lst:aspsaturation}, style=model, numbers=left]
map(G,X,v1) | map(G,X,v2) | map(G,X,v3) | map(G,X,v4) :- invar(X), negative(G). @\label{lstline:probspec}@
map(G,X,V) :- saturated(G), t_node(X), node(G,V).

saturated(G) :- t_edge(X,Y), map(G,X,V1), map(G,Y,V2), not edge(G,V1,V2), negative(G), invar(X), invar(Y).
saturated(G) :- map(G,X,V),  map(G,Y,V), X != Y, invar(X), invar(Y). // we cannot map two different template nodes to the same 

neg_homowith(G) :- not saturated(G), negative(G).

negative_count(N) :- N = #count{G:neg_homowith(G)}.
:- negative_count(N), N > 1.

\end{lstlisting}

\begin{lstlisting}[caption=ASP Canonicity template-based check, style=model]
iso(X,x1) | iso(X,x2) | iso(X,x3) | iso(X,x4) :- invar(X).

candidate_var(X) :- iso(_,X).

%not iso!
iso_saturated :- invar(X1), invar(X2), iso(X1,V1), iso(X2,V2),     t_edge(V1,V2), not t_edge(X1,X2). 
iso_saturated :- invar(X1), invar(X2), iso(X1,V1), iso(X2,V2), not t_edge(V1,V2),     t_edge(X1,X2).

iso(X,V) :- invar(X), t_node(V), iso_saturated.

d1(X) :-     invar(X), not candidate_var(X). 
d2(X) :- not invar(X),     candidate_var(X).

not_equal :- d1(X). % check that in fact candidate is different from the pattern itself
not_equal :- d2(X). % check that in fact candidate is different from the pattern itself

iso_saturated :- not not_equal. % should not be completely equal

min_d1(N) :- N = #min{ X: d1(X) }, not iso_saturated.
min_d2(N) :- N = #min{ X: d2(X) }, not iso_saturated.

iso_saturated :- min_d1(N1), min_d2(N2), N1 > N2.
\end{lstlisting}

\begin{lstlisting}[caption=ASP auxilary predicates, style=model]
%selects subpattern

t_path(X,Y) :- t_edge(X,Y), invar(X), invar(Y).
t_path(X,Y) :- t_edge(X,Z), t_path(Z,Y), invar(X).

:- invar(X), invar(Y), not t_path(X,Y).

0 { invar(X) } 1 :- t_node(X).
% auxilary constraints


edge(G,Y,X) :- edge(G,X,Y).
t_edge(Y,X) :- t_edge(X,Y).
node(G,Y)   :- edge(G,Y,_).
t_node(X)   :- t_edge(X,_).
\end{lstlisting}

\begin{lstlisting}[caption=ASP canonicity previous solution isomorphism check, style=model]
iso(s1,X,x1) | iso(s1,X,x2) :- invar(X).
iso(s2,X,x2) | iso(s2,X,x3) :- invar(X).

candidate_var(G,X) :- iso(G,_,X).

iso_saturated(G) :- invar(X1), invar(X2), iso(G,X1,V1), iso(G,X2,V2),     t_edge(V1,V2), not t_edge(X1,X2). 
iso_saturated(G) :- invar(X1), invar(X2), iso(G,X1,V1), iso(G,X2,V2), not t_edge(V1,V2),     t_edge(X1,X2). 
iso_saturatea(G) :- not equal(G), iso(G,_,_). 

iso(G,X,V) :- invar(X), t_node(V), iso_saturated(G).

:- not iso_saturated(G), iso(G,_,_).

d1(G,X) :-     invar(X), not candidate_var(G,X), iso(G,_,_).
d2(G,X) :- not invar(X),     candidate_var(G,X).

not_equal(G) :- d1(G,X). % check that in fact candidate is different from the pattern itself
not_equal(G) :- d2(G,X). % check that in fact candidate is different from the pattern itself

equal(G) :- not not_equal(G), iso(G,_,_).

\end{lstlisting}

\begin{lstlisting}[caption=ProB specification (without dataset), style=model]
MACHINE Knowledge
INCLUDES Dataset
SETS
  /* Two predefined sets exist, the vertices that the template and pattern can connect, and the labels.
   * The labels are already defined within Dataset.mch
   */
  Vertices = {x1,x2,x3,x4,x5,x6,x7,x8}
CONSTANTS
  /* The template and our pattern are the constants.
   *  * Template is given
   *  * Patterns is a set that must be found
   */
  Template,
  Patterns
DEFINITIONS

  SET_PREF_TIME_OUT == 70000; SET_PREF_MAX_INITIALISATIONS == 1;

  /* The (most general, i.e. ternary) definition of homomorphism. Note ' is the property accessor for records*/
  homomorph_with(FromGraph, iso, ToGraph) == (
    iso : Vertices >-> dom(ToGraph'LABEL) &
    !x.( x:Vertices => FromGraph'LABEL(x) = ToGraph'LABEL(iso(x))) &
    !(x,y).( x|->y : FromGraph'EDGES
         => iso(x)|->iso(y) : ToGraph'EDGES)
  );

  /* The (most general, i.e. ternary) definition of isomorphism*/
  isomorphic(FirstGraph, iso, SecondGraph) == (  
    #(V1,V2).(
    vertices(FirstGraph'EDGES, V1) &
    vertices(SecondGraph'EDGES, V2) &
    iso : V1 >->> V2 &
    !x.( x:V1 => FirstGraph'LABEL(x) = SecondGraph'LABEL(iso(x))) &
    !(x,y).( x|->y: FirstGraph'EDGES
        => iso(x)|->iso(y) : SecondGraph'EDGES) &
    !(x,y).( x|->y: SecondGraph'EDGES
        => iso~(x)|->iso~(y) : FirstGraph'EDGES)
    )
  );
  
  vertices(EdgeRelation, Vertices) == (
    Vertices = dom(EdgeRelation)  \/ ran(EdgeRelation)
  )

PROPERTIES

  /*This is our given template*/
  Template = {(x1,x2),(x2,x3),(x3,x4),(x4,x5),(x5,x6),(x6,x7),(x7,x8)} &

  /*Typing our Patterns set. It's a set of records (struct-type) with label a total function and edges a relation */
  Patterns : POW(struct(LABEL:Vertices-->Labels, EDGES:Vertices<->Vertices)) &
  /*Derived type: POW(struct(EDGES:POW(Vertices*Vertices), LABEL:POW(Vertices*Labels)))*/

  /*A single small test, this is not used anymore but is useful to check edits*/
  /* #isop.(homomorph_with(rec(LABEL:{(x1,a),(x2,b),(x3,a),(x4,a),(x5,a),(x6,a),(x7,a),(x8,a)}, EDGES:{(x1,x2),(x2,x3)}), isop, rec(LABEL:{(1,a),(2,a),(3,b),(4,a),(5,a),(6,a),(7,a),(8,a)}, EDGES:{(1,2),(2,3),(3,4)},SIGN:"POS"))) &*/

  /* Feed the pattern set with one specific pattern already */
  rec(LABEL:{(x1,a),(x2,b),(x3,a),(x4,a),(x5,a),(x6,a),(x7,a),(x8,a)}, EDGES:{(x1,x2),(x2,x3)}) : Patterns &

  /* Requirements on patterns:
   *  * The pattern is a subgraph of the template
   *  * The number of homomorphisms with positive graphs is great enough (at least-requirement)
   *  * The number of homomorphisms with negative graphs is small enough (at most-requirement)
   *  * No two patterns in the Patterns set are isomorphic
   */
  !pattern.(pattern:Patterns => pattern'EDGES <: Template) &
  !pattern.(pattern:Patterns => card({p|p:graphs & p'SIGN="POS" & #isop.(homomorph_with(pattern, isop, p))}) >= 1) &
  !pattern.(pattern:Patterns => card({p|p:graphs & p'SIGN="NEG" & #isop.(homomorph_with(pattern, isop, p))}) <= 0) &
  !(p1,p2).(p1:Patterns & p2:Patterns & p1 /= p2 => not (#iso.(isomorphic(p1, iso, p2)))) &

  #iso.(homomorph_with(rec(EDGES:{(x1|->x2)},LABEL:{(x1|->a),(x2|->a),(x3|->a),(x4|->a),(x5|->a),(x6|->a),(x7|->a),(x8|->a)}),iso,rec(EDGES:{(x1|->x2),(x3|->x4)},LABEL:{(x1|->a),(x2|->a),(x3|->a),(x4|->a),(x5|->a),(x6|->a),(x7|->a),(x8|->a)}))) & 

  /* We look for at least n patterns */
  card(Patterns) = 6 &

  1=1
OPERATIONS
 Pat(pattern) = SELECT pattern:Patterns THEN skip END
END
\end{lstlisting}

\end{document}